\RequirePackage{fix-cm}
\documentclass{svjour3}                     
\smartqed  
\usepackage{graphicx}
\usepackage{amsmath}
\usepackage{epstopdf}
\newcommand{\christoffelM}[3]{\ensuremath{\begin{pmatrix}\multicolumn{2}{c}{#1}&\\#2&#3\end{pmatrix}}}

\begin{document}

\title{Can negative~mass be considered in General~Relativity~?
}
\subtitle{}


\author{Jean-Pierre Petit         \and
        Gilles d’Agostini
}


\institute{F. Author \at
              \email{jppetit1937@yahoo.fr}           
           \and
           Gilles d’Agostini \at
              \email{dagostini.gilles@laposte.net}
}

\date{Received: date / Accepted: date}

\maketitle

\begin{abstract}
We show, through Newtonian approximation, that shifting to a bimetric model of the Universe based on a suitable system of coupled field equations, removes the preposterous runaway effect and gives different interaction laws, between positive and negative masses, that bring new insights into the alternative VLS interpretations previously proposed by several authors, and strengthening their assumptions. 
\end{abstract}

\section{Introduction}
\label{intro}
Scientists have been trying long ago to mix positive and negative masses to compose a new cosmological model. The lacunar structure of the Universe at large scales early suggested that some negative matter could have formed clumps, repelling positive matter into remnant places shaped like joint bubbles. 
Such a bubble structure was suggested in 1986 by De Lapparent et al. \cite{RefDelapparent} then by Kirshner et al.  \cite{RefKirshner} in 1986, Rood \cite{RefRood} in 1988 and Geller et al.  \cite{RefGeller} in 1989. The idea of repulsive structures, shaping positive matter into lacunar structures was progressively developped by El-Ad et al.   \cite{RefElAd1,RefElAd2,RefElAd3}.
\\
The possibility of gravitational repulsion was explicitely evoked by Piran \cite{RefPiran} in 1997, a paper in which he suggested that initial tiny perturbations in the primordial density field, “acting as if they had a negative mass”, could expand and play a role in the VLS (Very Large Scale) structure. He further theorized, with no justification however, that negative mass could own a negative active mass (from Bondi’s definition \cite{RefBondi}), but positive inertial and passive masses, so that two negative masses would mutually attract. Hence, negative material could form clusters while positive and negative masses would repel each other. He developped a Void Finder Algorithm allowing to identify voids in redshift surveys and to measure their size and “underdensity”. Interesting three-dimensional views of voids, based on the IRAS survey, are shown in his paper \cite{RefPiran}.
\\
2D-simulations involving the same interaction schema had been performed by Petit \cite{RefPetit1} in 1995, also giving a VLS lacunar structure.  More recently (July 2013) Izumi et al. \cite{RefKoki} suggested that the distortions observed on the images of distant galaxies, if considered as due to some negative gravitational lensing effect being radial instead of azimutal, would lead to a somewhat different mapping of the invisible matter.
\\
However, all these works contradicted Bondi’s result of 1957  \cite{RefBondi} based on General Relativity (GR), hence on the assumption that the Universe is a M4 manifold, coupled to a single metric g obeying Einstein’s equation. The Newtonian approximation gives such a metric a time independent quasi-Lorentzian form : 
\begin{equation}
g^{\mu \nu}\,=\,\eta ^{\mu \nu}\,+\,\varepsilon \gamma ^{\mu \nu}
\label{geps}
\end{equation}
\\
A subsequent development of the field equation into a series, combined to similar approximation of the equations of geodesics, provides Newton’s law and Poisson's equation. Bondi expressed the result as :
\begin{equation}
m^{(1)}_{i} \frac{d^{2}\overrightarrow{r_{1}}}{dt^{2}} = G m^{(1)}_{p} m^{(2)}_{a} \frac{(\overrightarrow{r_{2}}-\overrightarrow{r_{1}})}{|\overrightarrow{r_{2}}-\overrightarrow{r_{1}}|^{3}}
\label{gravbondi}
\end{equation}
accounting for the gravitational action of a mass (2) on a mass (1), and decided to introduce three kinds of masses : 
\\- inertial mass $m_{i}$
\\- passive gravitational mass $m_{p}$
\\- active gravitational mass $m_{a}$
\\
Because of the equivalence principle the first two are identical, so that Equ.~\ref{gravbondi} becomes : 
\begin{equation}
\frac{d^2\overrightarrow{r_1}}{dt^2} = G m^{(2)}_{a} \frac{(\overrightarrow{r_2}-\overrightarrow{r_1})}{|\overrightarrow{r_2}-\overrightarrow{r_1}|^3}
\label{grav}
\end{equation}
\\
In other words: according to GR, a positive mass attracts any kind of mass, whatever positive or negative. Conversely, a negative mass would repel any kind of mass. In particular, if one considers a couple of masses with equal absolute values but opposite signs, the negative mass escapes, while the positive mass runs after it, both being accelerated the same way. To increase such paradoxical phenomenon, because the masses have opposite signs, energy and momentum are conserved. 
\\
In addition, according to Bondi's analysis two negative masses mutually repel, so that negative matter cannot form clusters by gravitational accretion through Jeans’ instability. 
\\
Bondi tried to integrate such puzzling “runaway particles” in GR without any success. Only much later, in 1989, Bonnor \cite{RefBonnor} also published a paper devoted to negative mass. He considered the runaway phenomenon as “preposterous” and studied a universe only filled with negative masses, immediatly pointing out that this has nothing to do with physics and astrophysics, for such a medium cannot form stars and galaxies through gravitational instability. 
\\ \\
As a conclusion, in despite of interesting perspectives suggested by the previously mentioned references, GR bans any introduction of negative masses in the Universe, unless giving up the equivalence principle itself.  

\section{Bimetric Newtonian approximation}
A first tentative solution was suggested in 1994 by Petit \cite{RefPetit1,RefPetit2}, and later by Henry-Couannier \cite{RefCouannier}, then Hossenfelder \cite{RefHossenfelder}. All are based on a new description of the Universe, considered as a manifold M4 coupled to two metrics instead of a single one.
\\ \\
Basically, the first metric $g^{(+)}_{\mu\nu}$ refers to positive energy particles (and positive mass, if they have one), and $g^{(-)}_{\mu\nu}$ to negative energy particles (and negative mass, if they have one). Opposite energy particles cannot encounter, for they cruise on distinct geodesic families. Both families include null geodesics, with positive energy photons cruising on null geodesics built from metric $g^{(+)}_{\mu\nu}$, while negative energy photons cruise on null geodesics built  from metric $g^{(-)}_{\mu\nu}$.  It is assumed than positive structures, made of positive masses, can emit or absorb positive energy photons only, while negative structures, made of negative masses, can emit and absorb negative energy photons only. In short, on purely geometrical grounds, negative mass structures would be invisible for a positive mass observer (and vice-versa).
\\ \\
First, let us consider a system of coupled field equations with which we intend to describe a time-independant configuration, or quasi time-independant. The study of a time dependant solution, based on a modified system will be the subject of a next publication.
\begin{eqnarray}
R^{(+)}_{\mu\nu} - \frac{1}{2}R^{(+)}g^{(+)}_{\mu\nu} = +\chi \left(T^{(+)}_{\mu\nu}+T^{(-)}_{\mu\nu}\right) 
\nonumber \\\nonumber  \\
R^{(-)}_{\mu\nu} - \frac{1}{2}R^{(-)}g^{(-)}_{\mu\nu} = -\chi \left(T^{(+)}_{\mu\nu}+T^{(-)}_{\mu\nu}\right) 
\label{riccipm}
\end{eqnarray}
\\
We find the Ricci tensors built from metrics  $g^{(+)}_{\mu\nu}$ and $g^{(-)}_{\mu\nu}$. The energy-momentum tensors $T^{(+)}_{\mu\nu}$ and $T^{(-)}_{\mu\nu}$, in the simple case of non interacting incoherent matters, can be expressed in term of the scalar proper density fields $\rho^{(+)}$ and $\rho^{(-)}$ and four-velocity flows $u^{\mu}(x)=\frac{dx^{\mu}}{ds}$ by : $T^{\mu\nu} = \rho(x) u^{\mu}(x) u^{\nu}(x)$. 
\\\\
In the Newtonian approximation, for both species, the velocities are small with respect to the speed of light and the energy densities low. The energy-momentum tensors for a field of matter can be writen,in this case, as :
\begin{eqnarray}
{T^{(+)}}^{\mu}_{\nu}=\rho^{(+)}c^{2}
\begin{pmatrix}
\;1&\;0&\;0&\;0\;\\ 
\;0&\;0&\;0&\;0\;\\
\;0&\;0&\;0&\;0\;\\
\;0&\;0&\;0&\;0\;
\end{pmatrix}
\qquad\qquad
{T^{(-)}}^{\mu}_{\nu}= \rho^{(-)}c^{2}
\begin{pmatrix} 
\;1&\;0&\;0&\;0\;\\ 
\;0&\;0&\;0&\;0\;\\
\;0&\;0&\;0&\;0\;\\
\;0&\;0&\;0&\;0\;
\end{pmatrix} 
\label{tmixtem}
\end{eqnarray}
with $\rho^{(+)}>0$ \; and \; $\rho^{(-)}<0$ 
\\
\\
In the Newtonian approximation, the system (\ref{riccipm}) admits, for  $\rho^{(+)}+\rho^{(-)}=0$ a couple of Lorentzian solutions $\eta^{(+)}$ and  $\eta^{(-)}$ : 
\begin{equation}
\eta^{(+)}_{\mu\nu}=\eta^{(-)}_{\mu\nu}=
\begin{pmatrix}
\;1&\;\;0&\;\;0&\;\;0\;\\ 
\;0&-1&\;\;0&\;\;0\;\\
\;0&\;\;0&-1&\;\;0\;\\
\;0&\;\;0&\;\;0&-1\;
\end{pmatrix}
\equiv \eta_{\mu\nu}
\end{equation}
\\
Let us consider two time-independent small perturbations with finite extents, and write the metrics and the mass densities as :
\begin{eqnarray}
g^{(+)}_{\mu \nu}\,=\,\eta_{\mu \nu}\,+\,\varepsilon \gamma ^{(+)}_{\mu \nu}
\qquad\qquad
g^{(-)}_{\mu \nu}\,=\,\eta_{\mu \nu}\,+\,\varepsilon \gamma ^{(-)}_{\mu \nu}
\label{gepspm}
\end{eqnarray}
\begin{eqnarray}
\rho^{(+)}\,=\,\rho^{(+)}_{(0)}\,+\,\varepsilon \delta\rho^{(+)}
\qquad\qquad
\rho^{(-)}\,=\,\rho^{(-)}_{(0)}\,+\,\varepsilon \delta\rho^{(-)}
\label{gepspm}
\end{eqnarray}
with $\rho^{(+)}_{(0)}+\rho^{(-)}_{(0)}=0$. The mass densities are not supposed individually small.
\\ \\
We will now consider only the firt order perturbation terms. 
\\ \\
Let us introduce the Laue scalars~:
\begin{eqnarray}
T^{(+)}~\equiv~{T^{(+)}}^{\mu}_{\mu}~\simeq\rho^{(+)}c^{2}
\qquad\qquad
T^{(-)}~\equiv~{T^{(-)}}^{\mu}_{\mu}~\simeq\rho^{(-)}c^{2}\label{tdlaue}
\end{eqnarray}
in the system (\ref{riccipm}) which becomes :
\begin{eqnarray}
R^{(+)}_{\mu\nu} \simeq +\chi \left(T^{(+)}_{\mu\nu}+T^{(-)}_{\mu\nu}\right) - \frac{1}{2}\eta_{\mu\nu} \chi \left(T^{(+)}+T^{(-)}\right) 
\nonumber \\
R^{(-)}_{\mu\nu} \simeq -\chi \left(T^{(+)}_{\mu\nu}+T^{(-)}_{\mu\nu}\right)  + \frac{1}{2}\eta_{\mu\nu} \chi \left(T^{(+)}+T^{(-)}\right) 
\label{riccipm2}
\end{eqnarray}
\\
and then :
\begin{eqnarray}
R^{(+)}_{\mu\nu} = + \frac{\chi\varepsilon}{2}\left(\delta\rho^{(+)}+\delta\rho^{(-)}\right)c^2\delta_{\mu\nu}
\nonumber \\
R^{(-)}_{\mu\nu} = - \frac{\chi\varepsilon}{2}\left(\delta\rho^{(+)}+\delta\rho^{(-)}\right)c^2\delta_{\mu\nu}
\label{riccipm3}
\end{eqnarray}
\\
\\
Using the classical method and notations of \cite{RefABS}, the Ricci tensors are developed into series: 
\begin{eqnarray}
 R^{(+)}_{\mu\nu} \cong \left[\frac{1}{2}Ln(-g^{(+)})\right]^{(+)}_{|\mu|\nu} -\christoffelM{\alpha}{\mu}{\nu}^{(+)}_{|\alpha}  
\nonumber \\\nonumber \\
 R^{(-)}_{\mu\nu} \cong \left[\frac{1}{2}Ln(-g^{(-)})\right]^{(-)}_{|\mu|\nu} -\christoffelM{\alpha}{\mu}{\nu}^{(-)}_{|\alpha}  
\label{xxx} 
\end{eqnarray}
\\
Thus the approximate field equations may be expressend as :
\begin{eqnarray}
 \left[\frac{1}{2}Ln(-g^{(+)})\right]^{(+)}_{|\mu|\nu} -\christoffelM{\alpha}{\mu}{\nu}^{(+)}_{|\alpha} = + \frac{\chi\varepsilon}{2}\left(\delta\rho^{(+)}+\delta\rho^{(-)}\right)c^2\delta_{\mu\nu}
\nonumber \\\nonumber \\
\left[\frac{1}{2}Ln(-g^{(-)})\right]^{(-)}_{|\mu|\nu} -\christoffelM{\alpha}{\mu}{\nu}^{(-)}_{|\alpha} = - \frac{\chi\varepsilon}{2}\left(\delta\rho^{(+)}+\delta\rho^{(-)}\right)c^2\delta_{\mu\nu} 
\label{xxx} 
\end{eqnarray}
\\
Consider the case $\mu=\nu=0$,  with the time-independance hypothesis (index 0 denoting the time-marker), we get :
\begin{eqnarray}
\varepsilon \sum_{i=1}^{3}\left[\gamma^{(+)}_{0 0}\right]_{|\i|\i} = \varepsilon\left(\frac{\partial^2\gamma^{(+)}_{0 0}}{\partial x^2} + \frac{\partial^2\gamma^{(+)}_{0 0}}{\partial y^2} + \frac{\partial^2\gamma^{(+)}_{0 0}}{\partial z^2}\right)
= -\chi\varepsilon\left(\delta\rho^{(+)}+\delta\rho^{(-)}\right) c^{2}\qquad
\nonumber \\
\nonumber \\
\varepsilon \sum_{i=1}^{3}\left[\gamma^{(-)}_{0 0}\right]_{|\i|\i} = \varepsilon\left(\frac{\partial^2\gamma^{(-)}_{0 0}}{\partial x^2} + \frac{\partial^2\gamma^{(-)}_{0 0}}{\partial y^2} + \frac{\partial^2\gamma^{(-)}_{0 0}}{\partial z^2}\right)
= +\chi\varepsilon\left(\delta\rho^{(+)}+\delta\rho^{(-)}\right) c^{2}\qquad
\label{laplap}
\end{eqnarray}
\\
If we define the potentials as follows, the equations (\ref{laplap}) may be considered as a bimetric extension of the clasical Poisson equation~:
\begin{equation}
\varphi^{(+)}=\frac{c^2}{2} \varepsilon \gamma^{(+)}_{00}
\;\;\;\;\;\;\;\;\;
\varphi^{(-)}=\frac{c^2}{2} \varepsilon \gamma^{(-)}_{00}
\label{phipm}
\end{equation}
The constant $\chi$ is determined by considering a portion of space where $\rho^{(-)}=0$. The identification to classical Poisson's equation gives :
\begin{equation}
\chi=-\frac{8 \pi G}{c^4}
\label{xxx}
\end{equation}
\\
Then, for a bimetric system, we get an extended Poisson's equation~: 
\begin{eqnarray}
\triangle\varphi^{(+)} &=& + 4 \pi G \left(\rho^{(+)}+\rho^{(-)}\right)
\nonumber \\
\nonumber \\
\triangle\varphi^{(-)} &=& - 4 \pi G \left(\rho^{(+)}+\rho^{(-)}\right) = - \triangle\varphi^{(+)} 
\label{poissonpm}
\end{eqnarray}
and for the potentials :
\begin{eqnarray}
\varphi^{(+)}(\textbf{x}) &=& -G \int \dfrac{\left(\;\rho^{(+)}(\textbf{x'})+\rho^{(-)}(\textbf{x'})\;\right)}{|\textbf{x}-\textbf{x'}|}{d^3\textbf{x}} \equiv \varphi(\textbf{x})
\nonumber \\
\nonumber \\
\varphi^{(-)}(\textbf{x}) &=& - \varphi^{(+)}(\textbf{x}) = -\varphi(\textbf{x})
\label{potentielpm}
\end{eqnarray}

\section{Gravity as a Bimetric Phenomenon}
Let us write the two metrics :
\begin{eqnarray}
(ds^{(+)})^2 = c^2 dt^2 - (dx^1)^2 - (dx^2)^2 - (dx^3)^2 + \varepsilon \gamma^{(+)}_{\mu\nu} dx^{\mu} dx^{\nu}
\nonumber \\
\nonumber \\
(ds^{(-)})^2 = c^2 dt^2 - (dx^1)^2 - (dx^2)^2 - (dx^3)^2 + \varepsilon \gamma^{(-)}_{\mu\nu} dx^{\mu} dx^{\nu}
\label{xxx}
\end{eqnarray}
and the corresponding differential equations of the geodesics~:
\begin{eqnarray}
\frac{d^2x^{\alpha}}{ds^{(+)2}} + \christoffelM{\alpha}{\eta}{\tau}^{(+)}\frac{dx^{\eta}}{ds^{(+)}}\frac{dx^{\tau}}{ds^{(+)}}=0
\nonumber \\
\nonumber \\
\frac{d^2x^{\alpha}}{ds^{(-)2}} + \christoffelM{\alpha}{\eta}{\tau}^{(-)}\frac{dx^{\eta}}{ds^{(-)}}\frac{dx^{\tau}}{ds^{(-)}}=0
\label{geodpm} 
\end{eqnarray}
\\
Hence, to the first order in $\varepsilon$ and $\beta=\sqrt{\frac{(dx^1)^2 + (dx^2)^2 + (dx^3)^2}{dt^2}}\equiv(\dfrac{v}{c})$, equations (\ref{geodpm}) becomes~:
\begin{eqnarray}
\frac{d^2x^{\alpha}}{ds^{(+)2}} + \christoffelM{\alpha}{0}{0}^{(+)}\left(\frac{dx^0}{ds^{(+)}}\right)^2=0
\nonumber \\
\nonumber \\
\frac{d^2x^{\alpha}}{ds^{(-)2}} + \christoffelM{\alpha}{0}{0}^{(-)}\left(\frac{dx^0}{ds^{(-)}}\right)^2=0
\label{geod2pm} 
\end{eqnarray}
\\and we have also~:
\begin{equation}
\left(\frac{ds^{(+)}}{dt}\right)^2 \cong \;c^2 (1+\varepsilon\gamma^{(+)}_{00})
\;\;\;\;\;\;\;\;
\left(\frac{ds^{(-)}}{dt}\right)^2 \cong \;c^2 (1+\varepsilon\gamma^{(-)}_{00})
\label{xxx}
\end{equation}
\\
After some calculation we derive the equations for two classes of geodesics :
\begin{eqnarray}
\frac{d^2x^i}{dt^2} = -\frac{c^2}{2} \varepsilon \frac{\partial\gamma^{(+)}_{00}}{\partial x^i} = -\frac{\partial\phi^{(+)}}{\partial x^i}
\nonumber \\
\nonumber \\
\frac{d^2x^i}{dt^2} = -\frac{c^2}{2} \varepsilon \frac{\partial\gamma^{(-)}_{00}}{\partial x^i} = -\frac{\partial\phi^{(-)}}{\partial x^i}
\label{xxx}
\end{eqnarray}
\\
\vbox{
from which we get, with (\ref{poissonpm}) and (\ref{potentielpm}), the following laws of interaction : 
\begin{itemize}
\item Two positive masses attract each other through Newton’s law ;
\item Two negative masses attract each other through Newton’s law ;
\item Two masses, with opposite signs, repel each other through “anti-Newton’s” law.
\end{itemize}
}
The preposterous “runaway phenomenon” has disappeared. 

\section{Comparison with observational data}
If positive and negative masses repel each other, anywhere positive mass dominates, negative mass is almost absent, so that in the vicinity of the solar system $\rho^{(-)}\simeq 0$. Our coupled field equations become : 
\begin{equation}
R^{(+)}_{\mu\nu} - \frac{1}{2}R^{(+)}g^{(+)}_{\mu\nu} \simeq +\chi T^{(+)}_{\mu\nu}  
\label{einsteinp}
\end{equation}
\begin{equation}
R^{(-)}_{\mu\nu} - \frac{1}{2}R^{(-)}g^{(-)}_{\mu\nu} \simeq -\chi T^{(-)}_{\mu\nu}  
\label{einsteinm}
\end{equation}
\\
(\ref{einsteinp}) identifying to Einstein’s equation with a null cosmological constant, so that the bimetric model fits the classical local tests of GR.

\section{About Schwarzschild's solutions}
Theoreticians also noticed long ago that the parameter \textit{m} of Schwarzschild's solution was nothing but an integration constant, that may be chosen positive or negative. We will call the following metric “external Schwarschild solution”. 
\begin{equation}
ds^2=\left(1-\frac{2m}{r}\right)c^2 dt^2 - \frac{dr^2}{\left(1-\frac{2m}{r}\right)} - r^2\left(d\theta^2 + sin^2(\theta)d\phi^2\right)
\label{schwarp}
\end{equation}
\\
Similarly, in the “internal Schwarzschild solution” : 
\begin{eqnarray}
ds^2=\left[\frac{3}{2}\sqrt{1-\frac{r^2_0}{A}}-\frac{1}{2}\sqrt{1-\frac{r^2}{A}}\right]c^2 dt^2
-\frac{dr^2}{1-\frac{r^2}{A}} -r^2\left(d\theta^2 + sin^2(\theta)d\phi^2\right)
\nonumber \\
for \; r\leq r_0 \;\;  with \;\;  A=\frac{3c^2}{8\pi G \rho}\; \; \; 
\label{schwarm}
\end{eqnarray}
\\
$\rho$ can be chosen positive or negative, with the coupling condition : 
\begin{equation}
m = \frac{4 \pi G r^3_0}{c^2}\rho \qquad (\;\underline{positive}\;or\;\underline{negative}\;)
\label{xxx}
\end{equation}
\\
In GR, the internal Schwarzschild solution, describing the geometry inside a sphere filled  with constant matter density, was considered as some kind of mathematical complement to the external solution, since no particle would follow a geodesic inside. In the bimetric approach (\ref{schwarp}) and (\ref{schwarm}) provide the geodesics around clumps of negative mass, corresponding to the equations :
\begin{equation}
R^{(+)}_{\mu\nu} - \frac{1}{2}R^{(+)}g^{(+)}_{\mu\nu} \simeq -\chi T^{(-)}_{\mu\nu}  
\label{inducep}
\end{equation}
\begin{equation}
R^{(-)}_{\mu\nu} - \frac{1}{2}R^{(-)}g^{(+)}_{\mu\nu} \simeq +\chi T^{(-)}_{\mu\nu}  
\label{inducem}
\end{equation}
\\
Equation (\ref{inducep}) corresponds to an “induced geometry” in the positive sector, due to the presence of some negative mass. A positive energy photon can cross freely a negative mass. Considering positive energy photons, this give a negative lensing effect not only in the vicinity of a negative mass, but also inside it,  as firstly suggested in \cite{RefPetit1}. The present work supports the assumptions corresponding to reference  \cite{RefPiran}.

\section{Conclusion}
The classical General Relativity model bans the existence of negative mass in the Universe, while implying a puzzling phenomenon : the runaway effect. After GR, positive mass attracts anything and negative mass repels anything, so that if two masses with opposite signs encounter, the positive one escapes and the negative one runs after it since both experience the same acceleration, while energy and momentum are conserved. G.W. Bonnor qualified such phenomenon as preposterous. However, choosing other interaction laws provides interesting challenging interpretations of the VLS as shown in the present work,  based on a new bimetric model of the Universe supporting the various assumptions of previous investigators and opening the road towards interesting research. 




\end{document}